\documentstyle[aps,epsf,rotate,multicol]{revtex}
\begin{document}
\draft

\title{Symmetry Breaking in Stock Demand}

\author{Vasiliki Plerou, Parameswaran Gopikrishnan, and 
H. Eugene Stanley}

\address{Center for Polymer Studies and Department of Physics,
Boston University, Boston, Massachusetts 02215.\\}

\date{\today}

\maketitle

\begin{abstract} 

Scale-free distributions and correlation functions found in financial
data are reminiscent of the scale invariance of physical observables
in the vicinity of a critical point. Here, we present empirical
evidence for a transition phenomenon, accompanied by a symmetry
breaking, in the investors' demand for stocks.  We study the volume
imbalance $\Omega$ --- difference between the number of shares traded
in buyer-initiated and seller-initiated trades in a time interval
$\Delta t$ --- conditioned on $\Sigma$ which is defined as the local
first moment of $\Omega$ in $\Delta t$. We find that the conditional
distribution $P(\Omega \vert \Sigma)$ undergoes a qualitative change
in behavior as $\Sigma$ increases beyond a critical threshold
$\Sigma_c$. For $\Sigma <\Sigma_c$, $P(\Omega\vert\Sigma)$ displays a
maximum at $\Omega=0$, i.e., trades in $\Delta t$ are equally likely
to be buyer initiated or seller initiated. For $\Sigma > \Sigma_c$,
$\Omega=0$ becomes a local minimum and two new maxima $\Omega_{+}$ and
$\Omega_{-}$ appear at non-zero values of $\Omega$, i.e., trades in
$\Delta t$ are either predominantly buyer initiated or predominantly
seller initiated. We interpret these results using a Langevin equation
with multiplicative noise.

\end{abstract}
\begin{multicols}{2}

Phase transitions of systems from an `ordered' phase into a
`disordered' phase are closely linked to symmetry breaking.  For
example, in an Ising ferromagnet above its critical temperature, the
most probable state (zero net magnetization) possesses the symmetry
that leaves it invariant under the flipping of each spin. For
temperatures below the critical value, this symmetry is broken, and
there is a preferred direction.

Here, we present evidence for an analogous transition phenomenon in a
financial
context~\cite{Takayasu02,Takayasu99,Bouchaud01,Zhang,Mantegna99,Bouchaud00,Farmer99}.
Specifically, we study the statistical properties of investors'
demand for stocks --- quantified as the imbalance in the number of
shares transacted by buyers and sellers over a time interval $\Delta
t$. We analyze the probability distribution of demand, conditioned on
its local ``noise'' intensity (a variance-like parameter $\Sigma$
defined below). We find that for intensities smaller than a critical
value $\Sigma_c$, the most probable value of demand is approximately
zero --- neither buying nor selling behavior dominates. For
intensities larger than that critical value ($\Sigma > \Sigma_c$), two
most probable values emerge that are symmetric around zero demand,
corresponding to two distinct ``phases'' --- excess demand and excess
supply~\cite{Takayasu99}. Under such conditions, the market behavior
is either mainly-buying or mainly-selling, spending almost equal
amount of time in each state.  In other words, exchanging every
``buy'' with a ``sell'' gives the same state below the critical noise
intensity, whereas above this threshold, the symmetry of this exchange
is broken.

In classic critical phenomena, the qualitative change in behavior
accompanying a phase transition can be formalized in terms of the
extrema of a phenomenological potential, or equivalently in terms of
the extrema of the corresponding probability
distributions~\cite{Halperin77,Stanley71}.  We first follow the latter
approach and study the behavior of the probability distribution of
demand.  Using transactions and quotes data for the 116 most-actively
traded stocks~\cite{TAQrec}, we quantify the demand for a stock by
calculating the ``volume imbalance'' over a time interval $\Delta t$,
defined to be the difference between $Q_{\rm B}$, the number of shares
traded in buyer-initiated trades, and $Q_{\rm S}$, the number of shares
traded in seller-initiated trades in $\Delta
t$~\cite{CLM,LeeReady91,LeeReady91note,Plerou01,Hasbrouck88,Farmer98},
\begin{equation}
\Omega (t) \equiv Q_{\rm B} - Q_{\rm S} = \sum_{i=1}^N q_i a_i\,.
\label{defOmega}
\end{equation}
Here, the indicator $a_i=1$ for buyer-initiated trades (buy trades)
and $a_i=-1$ for seller-initiated trades (sell trades)~\cite{buysell},
$q_i$ is the number of shares traded in transaction $i$, and $N \equiv
N_{\Delta t}$ denotes the number of trades in $\Delta
t$~\cite{TAQrec,blockTrades}.  

Our analysis of the (unconditional) probability distribution
$P(\Omega)$ for each stock shows a single peak around
$\Omega=0$. Since previous work shows that the distribution of $q_i$
has divergent variance~\cite{Gopi00a,notedistQprime}, we quantify the
noise intensity by computing the `local
deviation'~\cite{controlparameter}, defined as the centered first
moment,
\begin{equation}
\Sigma (t) \equiv \langle \vert q_i a_i- \langle q_i a_i \rangle \vert \rangle \,,
\label{defsigma}
\end{equation}
where $\langle\dots\rangle$ denotes `local' expectation values
computed from all trades in the time interval $\Delta t$. Next, we
examine the behavior of the conditional distribution $P (\Omega \vert
\Sigma)$ of $\Omega$ for a given value of the local deviation $\Sigma$
for $\Delta t=15$ min, [Fig.~1(a)].  For small $\Sigma$, we find that
$P(\Omega \vert \Sigma)$ is {\it single peaked} displaying a maximum
at $\Omega=0$. When $\Sigma$ exceeds a critical threshold $\Sigma_c$,
the behavior of $P(\Omega \vert \Sigma)$ undergoes a qualitative
change, and is {\it double peaked} with two new maxima appearing at
non-zero values, $\Omega_{+}$ and $\Omega_{-}$, symmetric around
zero. Figure~1(a) also shows that the separation between the two
maxima increases with $\Sigma$.

This qualitative change in the behavior of $P(\Omega)$ implies that
for $\Sigma < \Sigma_c$, the most-probable value of demand is
approximately zero, and possesses the symmetry that leaves the most
probable value invariant under the operation $\Omega \rightarrow
-\Omega$, or at the microscopic (trade) level, under the operation
${\rm B}\rightarrow {\rm S}$ of changing every buyer-initiated
trade~B, to a seller-initiated trade~S. For $\Sigma > \Sigma_c$, the
two most probable values $\Omega_{\pm}$ are non-zero, and the $\Omega
\rightarrow -\Omega$ (${\rm B}\rightarrow {\rm S}$) symmetry is
broken. In other words, while for $\Sigma < \Sigma_c$ buy and sell
trades are equally probable in each time interval (zero demand), for
$\Sigma > \Sigma_c$, trades in each time interval are either mostly
buy trades (excess demand) or mostly sell trades (excess supply)
giving rise to non-zero values of $\Omega_{\pm}$. Identical results
can be obtained by conditioning $P(\Omega)$ on the total trade volume
in $\Delta t$,  $Q(t)\equiv Q_{\rm B} + Q_{\rm S}$. 

Our finding is analogous to phase transition phenomena in physical
systems, where the behavior of the system undergoes a qualitative
change at a critical threshold of a control parameter
$T$. In such systems, the change in behavior
can be quantified by an order parameter $\Psi$ which is identically
zero for values of $T$ below (or above as the case may be) a certain
critical value $T_{\rm c}$, and becomes nonzero as $T$ crosses $T_{\rm
c}$. In our problem, the ``order parameter'' $\Psi$, can be identified
by the location of the maxima $\Omega_{\pm}$ of $P
(\Omega)$. Figure~1(c) shows that the change in $\Psi$ as a function
of $\Sigma$ is described by
\begin{equation}  
\Psi (\Sigma) = \cases{ 0 & $[\Sigma < \Sigma_c]$ \cr 
\vert \Sigma - \Sigma_c \vert^{\beta} & $[\Sigma > \Sigma_c]$},
\end{equation}
with $\beta \approx 1$~\cite{notebetahalf}.

In the mean-field theory of critical phenomena (Landau-Ginzburg
theory), the qualitative change in behavior of the system is
attributed to the changes in symmetry of the underlying
potential~\cite{Halperin77,Stanley71}. In the following, we pursue an
analogous approach to understand our empirical results. Since the
transition behavior that we find occurs with change in noise
intensity, we follow an approach similar to those used to understand
non-equilibrium phase transition
phenomena~\cite{Horsthemke,Haken75,Haken78}. We start with expressing
the dynamics of $\Omega$ through a {\it deterministic} differential
equation,
\begin{mathletters}
\begin{equation}
d\Omega = h_{\lambda}(\Omega)\, dt\,,
\label{determinomega}
\end{equation} 
where $\lambda$ is a parameter quantifying the coupling of the system
to its environment. Letting $\lambda$ fluctuate randomly with noise
intensity $\sigma$, Eq.(\ref{determinomega}) becomes, in general, a
{\it stochastic} differential equation with multiplicative
noise. Since the form of $h_{\lambda}(\Omega)$ is not known, and since
it is not {\it a priori} clear if $\lambda$ is an observable, we
describe the dynamics of $\Omega$ through~\cite{Horsthemke}
\begin{equation}
d\Omega = u (\Omega) dt + \sigma\, v(\Omega) dW_t\,,
\label{defdomega}
\end{equation}
\end{mathletters}
where $u\equiv u(\Omega)$ is the drift term, $v\equiv v(\Omega)$
reflects the effects of multiplicative noise, $dW_t$ is the standard
Wiener differential satisfying $\langle dW_t\,dW_{t^{\prime}} \rangle
=\delta (t-t^{\prime}) dt$, and the parameter $\sigma$ quantifies the
intensity of the noise term~\cite{lambda,fractional}. The functions
$u$ and $v$ may estimated from the data as a ``local'' mean and
standard deviation of $d\Omega$~\cite{notediff}.  The Fokker-Planck
equation corresponding to Eq.~(\ref{defdomega}) which describes the
evolution of the transition probability density $\Pi
\equiv P(\Omega, t \vert \Omega (t=0))$ is
\begin{equation}
{\partial \Pi \over \partial t} = -{\partial \over \partial \Omega} [u
\Pi] + {\sigma^2 \over 2} {\partial^2 \over \partial \Omega^2} [v^2\Pi]\,.
\label{fpe}
\end{equation}

The stationary probability density $P(\Omega)$ from Eq.~(\ref{fpe}) is
\begin{equation}
P (\Omega) = {A \over v^2} \exp \left( {2\over \sigma^2} 
\int^{\Omega} {u \over v^2} dx \right)\,,
\label{fpesol}
\end{equation}
where $A$ is the normalization constant, and $P$ is assumed
to be normalizable. Equation~(\ref{fpesol}) can be rewritten as 
\begin{equation}
P (\Omega) = A\, \exp\left(-{V(\Omega)\over\sigma^2}\right)\,,
\label{fpesolpot}
\end{equation}
where the function $V (\Omega)$ takes the meaning of a
effective ``stochastic'' potential given by~\cite{Horsthemke}
\begin{equation}
V (\Omega) \equiv -\left[\int^{\Omega} {u \over v^2} dx
- \sigma^2 \ln v
\right] \,.
\label{defpot}
\end{equation}
Thus, the extrema of the probability density can be translated into
the extrema of the underlying stochastic potential. 

As the noise intensity of the environment ($\sigma$) changes, the
potential could change shape, acquire new minima, and consequently the
system could undergo drastic changes in behavior~\cite{noteCONTB}. Can
the transition behavior we find empirically for $P(\Omega \vert
\Sigma)$ be understood in this framework? To address this question, we
must examine the shape of the potential $V (\Omega)$ for
different values of $\Sigma$, which monotonically depends on
$\sigma$~\cite{noteSigma-sigma}.

In order to study $V(\Omega)$ for different $\sigma$ empirically, we
first extract the functions $u(\Omega)$ and $v(\Omega)$ and analyze
their behavior for different values of $\Sigma$. From
Eq.~(\ref{defdomega}), $u$ is the drift term, which for a given
$\Sigma$, can be estimated by computing the equal-time expectation
value of the change $\Delta
\Omega \equiv \Omega (t+\Delta t) - \Omega(t)$ for a given $\Omega$,
\begin{equation}
u \approx \langle \Delta \Omega \rangle_{(\Omega,\Sigma)}\,.
\label{Fest}
\end{equation}
Similarly, the product $\sigma v$ can be estimated from the ``local''
deviation
\begin{equation}
\sigma\, v \approx \langle \vert \Delta \Omega  - \langle \Delta \Omega 
\rangle_{(\Omega,\Sigma)} \vert \rangle_{(\Omega, \Sigma)}\,.
\label{Gest}
\end{equation}

Figures~\ref{FG}(a) and (b) show $u (\Omega)$ and $v (\Omega)$ for
three different values of $\Sigma$. Clearly, the functional form
$u(\Omega)$ does not vary with $\Sigma$ and is consistent with a
linear behavior for all $\Sigma$. Figure~\ref{FG}(b) shows that, for
small $\Sigma$, $\sigma\,v (\Omega)$ is approximately flat, whereas
for large $\Sigma$, $\sigma\,v$ acquires a marked `peak' around
$\Omega=0$. Except for the smallest $\Sigma$, the functional forms
$\sigma v (\Omega)$ for different $\Sigma$ seem to be consistent
within a multiplicative factor (related to
$\sigma$)~\cite{note-additivenoise}.

Next, we shall analyze the extrema of the stochastic potential $V
(\Omega)$ for different values of $\Sigma$. From
Eqs.~(\ref{fpesolpot})~and~(\ref{defpot}), the extrema of $P(\Omega)$
correspond to the roots of the function
\begin{equation}
F (\Omega) \equiv {dV(\Omega) \over d\Omega} = u -
\sigma^2 v {dv \over d \Omega}.
\label{defFmaxmin}
\end{equation}
Figure~\ref{min}(a) shows $ F (\Omega)$ for three different
$\Sigma$. For $\Sigma< \Sigma_c$, the function $F (\Omega)$
displays only one root at $\Omega=0$. Near $\Sigma = \Sigma_c$, $ F
(\Omega)$ displays an inflexion point at $\Omega=0$. For $\Sigma >
\Sigma_c$, the one root branches into three: $\Omega=0$, $\Omega_{+}$,
and $\Omega_{-}$.

Integrating the stochastic ``force'' $ F (\Omega)$, we next find the
potential $V (\Omega)$.  Figure~\ref{potential} shows the behavior of
$V(\Omega)$ with different $\Sigma$.  For $\Sigma<
\Sigma_c$, the potential has only one minimum at $\Omega=0$, which is
consistent with one maximum for $P(\Omega)$ which we find. For $\Sigma
\approx \Sigma_c$, the potential appears almost flat: the existing
minimum begins to change into a (unstable) local maximum and displays
an inflexion point at $\Sigma=0$. For $\Sigma > \Sigma_c$, the
potential displays two clear minima at $\Omega_{\pm}$ and a local
maximum at $\Omega=0$, consistent with the bimodal nature of the
distribution $P(\Omega)$ for $\Sigma > \Sigma_c$.

In summary, we investigate the dynamics of the demand $\Omega$ by
examining the distribution of volume imbalance $\Omega$ for changing
market conditions quantified by the local deviation $\Sigma$. We find
that the distribution $P(\Omega)$ is single peaked for $\Sigma$
smaller than a critical threshold $\Sigma_c$. For $\Sigma$ larger than
$\Sigma_c$, the distribution $P(\Omega)$ changes to a double-peaked
distribution. The analog of the order parameter $\Psi$ which describes
the above qualitative change in behavior is zero for $\Sigma <
\Sigma_c$ and behaves as $\sim \vert \Sigma - \Sigma_c
\vert^{\beta}$~for $\Sigma > \Sigma_c$, where $\beta \approx 1$.
We have also seen that the dynamics of demand can be understood in
terms of a `stochastic' potential which changes its behavior with
$\Sigma$. As $\Sigma$ crosses the critical threshold $\Sigma_c$, the
system undergoes a transition from a `disordered' state with most
probable demand equal to zero to an `ordered' state with two phases,
excess demand and excess supply.

We thank the National Science Foundation for support.

\begin{figure}
\narrowtext
\centerline{
\epsfysize=0.7\columnwidth{\rotate[r]{\epsfbox{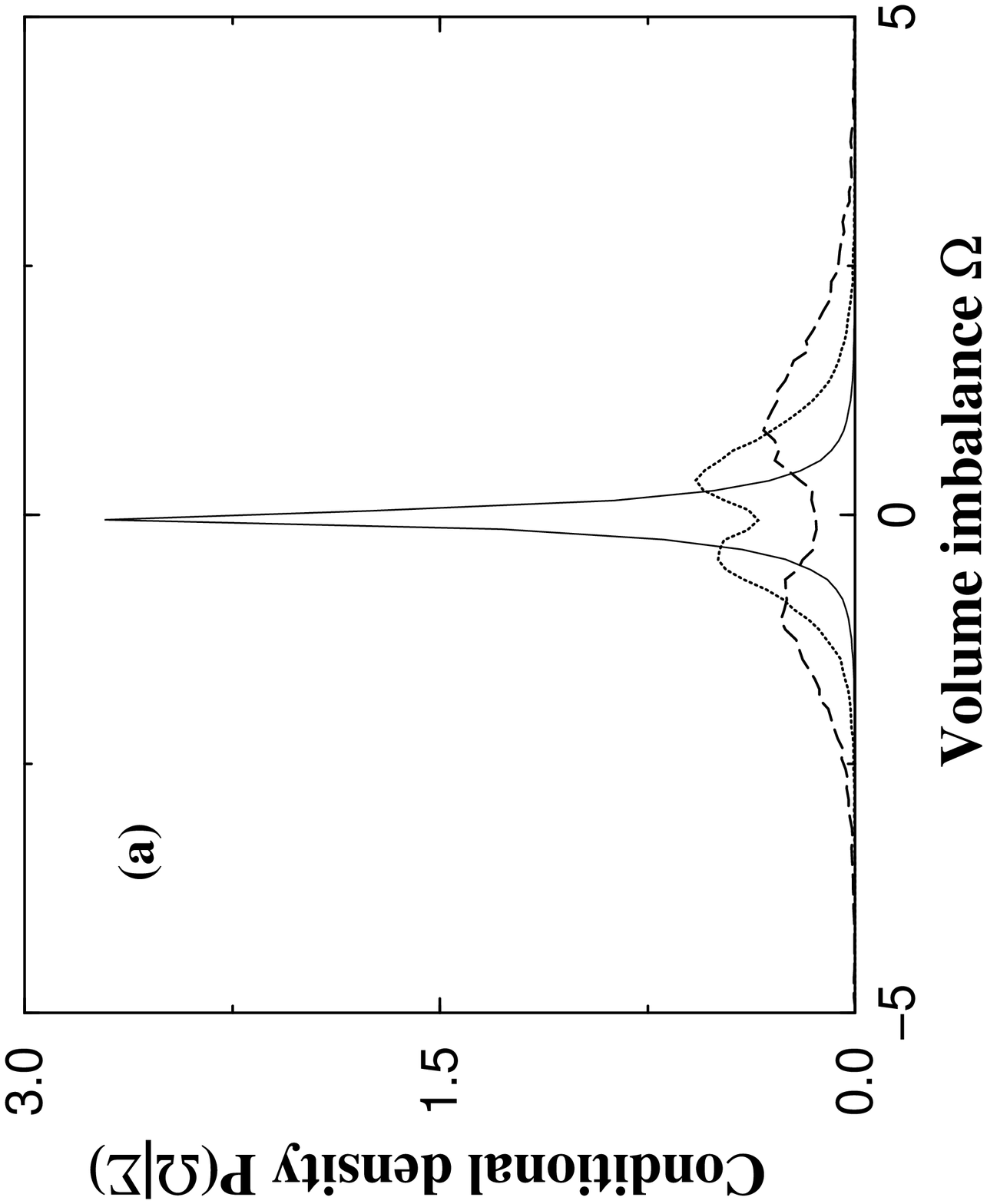}}}
}
\centerline{
\epsfysize=0.7\columnwidth{\rotate[r]{\epsfbox{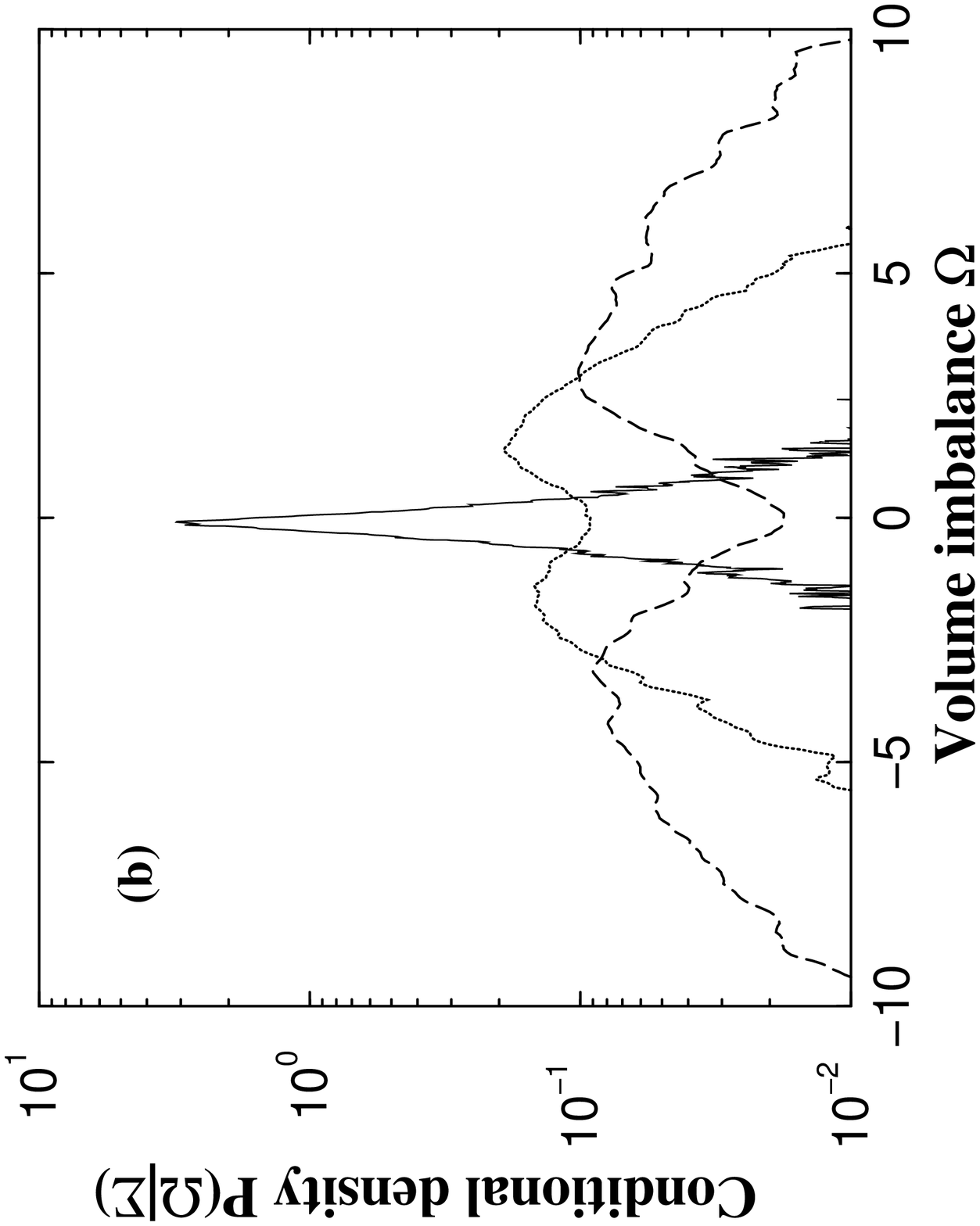}}}
}
\centerline{
\epsfysize=0.7\columnwidth{\rotate[r]{\epsfbox{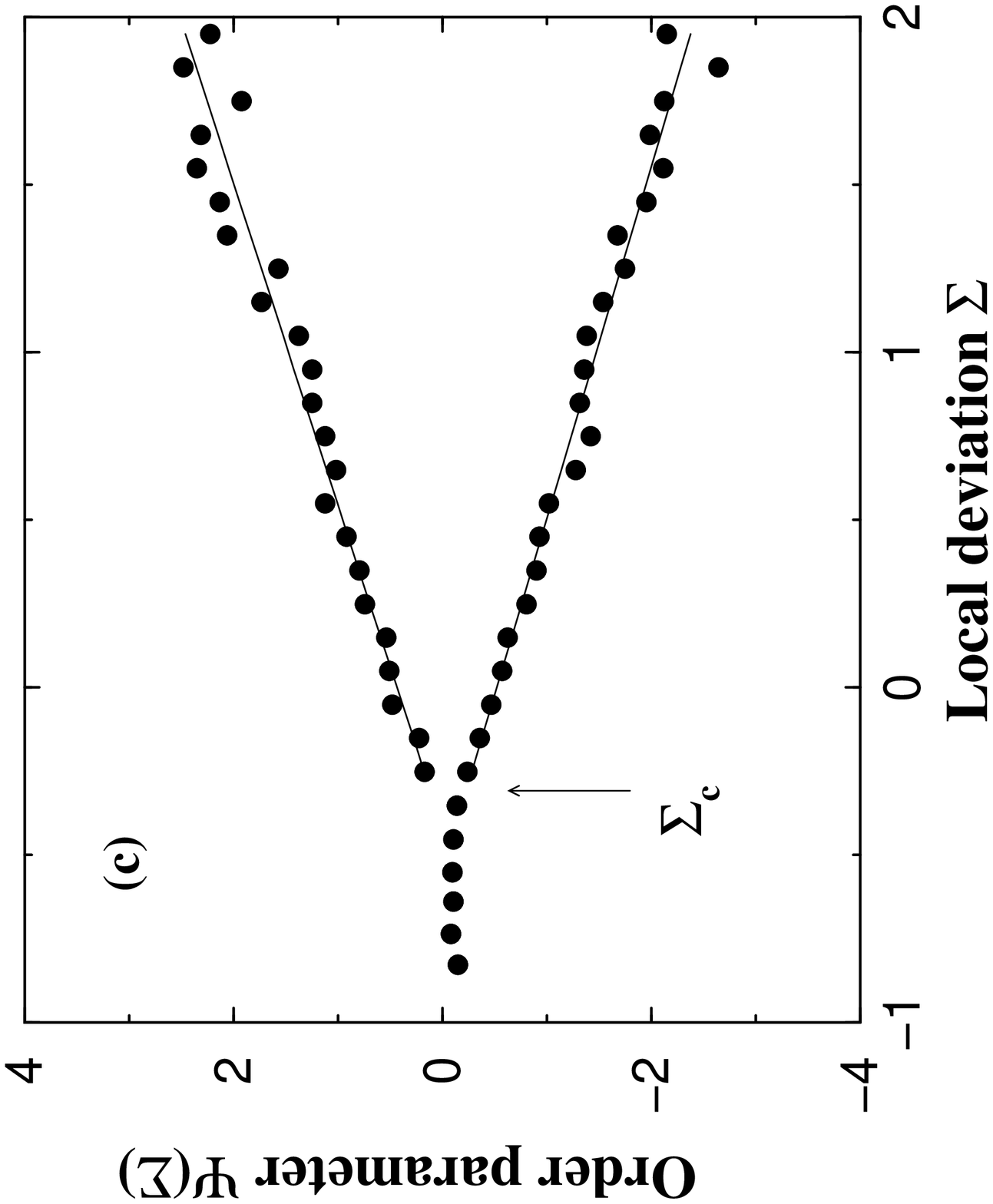}}}
}
\caption{(a) Conditional density  $P(\Omega \vert \Sigma)$ for
varying $\Sigma$. The distribution changes from a single-peaked
distribution for small $\Sigma$ (solid line), to a double-peaked
distribution for large $\Sigma$ (dashed line). (b) Same as (a) in
semi-logarithmic scale. (c) Order parameter, $\Psi$ (position of the
maxima of the distribution $P(\Omega \vert \Sigma)$), as a function of
$\Sigma$. For small $\Sigma$, $P(\Omega \vert
\Sigma)$ displays a single maximum whereas for large $\Sigma$ two
maxima appear. Both $\Sigma$ and $\Omega$ have been normalized to zero
mean and unit first moment. For a more accurate estimation of the
location of the extrema, all densities were computed using the density
estimator of~\protect\cite{Holy97}.  }
\label{qp_dens}
\end{figure}

\begin{figure}
\narrowtext
\centerline{
\epsfysize=0.7\columnwidth{\rotate[r]{\epsfbox{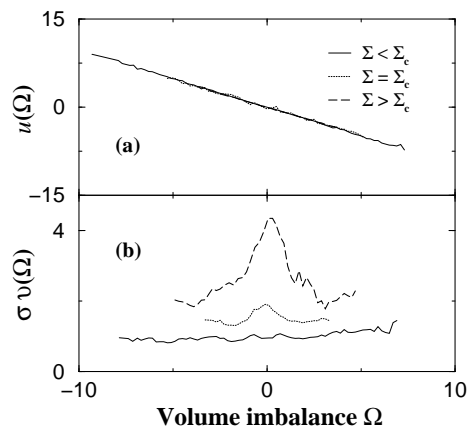}}}
}
\caption{(a) The drift part $u(\Omega)$  and (b) the variance part 
$\sigma\,v(\Omega)$ of Eq~(\protect\ref{defOmega}), estimated from the
data as the local mean and local deviation of $\Delta\Omega$. The
curves in~(b) have been shifted vertically for clarity.}
\label{FG}
\end{figure}

\begin{figure}
\narrowtext
\centerline{
\epsfysize=0.7\columnwidth{\rotate[r]{\epsfbox{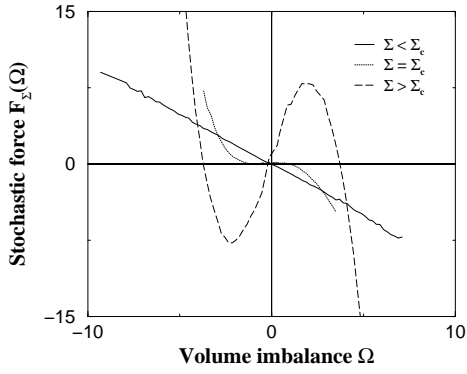}}}
}
\caption{Stochastic ``force'' $F(\Omega)$ for three
different values of $\Sigma$ calculated from
Eq.~(\protect\ref{defFmaxmin}).  The derivative $\sigma\,dv/d\Omega$
is calculated by first fitting the function $\sigma\,v$ by a third
order polynomial. For $\Sigma<
\Sigma_c$, the function $ F (\Omega)$ displays only one
zero at $\Omega=0$. For $\Sigma \approx \Sigma_c$, the existing root
starts to branch into three roots, and for $\Sigma > \Sigma_c$, we
find three roots: $\Omega=0$ and $\Omega=\Omega_{\pm}$.}
\label{min}
\end{figure}

\begin{figure}
\narrowtext
\centerline{
\epsfysize=0.7\columnwidth{\rotate[r]{\epsfbox{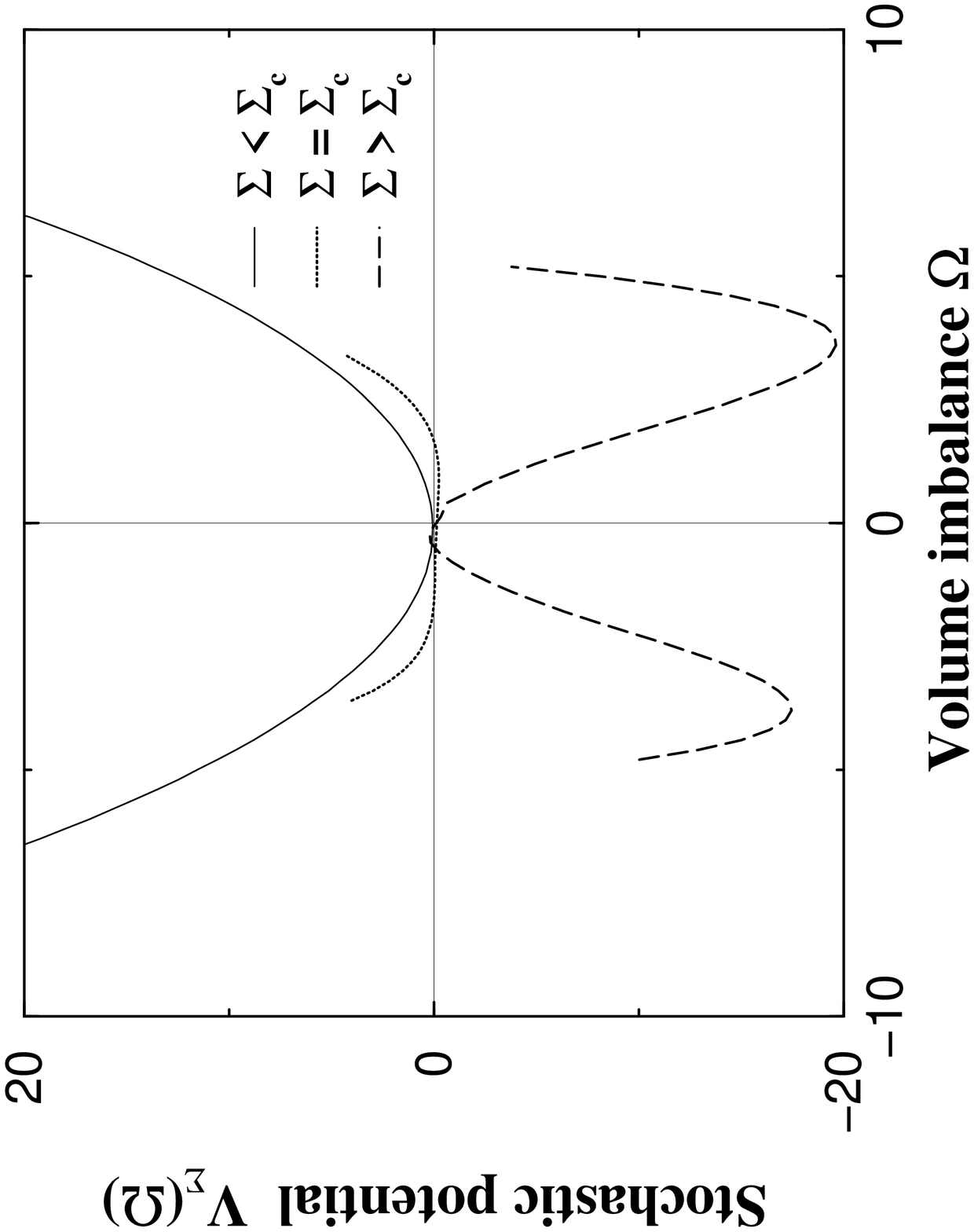}}}
}
\caption{Stochastic potential $V(\Omega)$ for 
$\Sigma < \Sigma_c$ shows one minimum at $\Omega=0$, whereas for
$\Sigma > \Sigma_c$ the potential has a maximum at $\Omega=0$ and two
new minima $\Omega_{\pm}$ appear. We compute $V_{\Sigma}(\Omega)$ by
integrating $F (\Omega)$ [Fig.~\protect\ref{min}]. The curves have
been shifted vertically for clarity.}
\label{potential}
\end{figure}
\end{multicols}


\begin{references}
\bibitem{blockTrades}
We disregard the first trade of each day as it is most often a block trade.
\bibitem{notedistQprime}
In Ref.~\protect\cite{Gopi00a} it is shown that $q_i$ have a power-law
  distribution with exponent $\zeta\approx 3/2$ which corresponds to a
  divergent second moment.
\bibitem{controlparameter}
We note that the results presented below are robust to using the `local
  deviation' of $\Omega$ ($=\Sigma N$) instead of $\Sigma$.
\bibitem{lambda}
In the literature, the diffusion term in Eq.~(\protect\ref{defdomega}) is
  attributed to fluctuations of a parameter $\lambda$ coupled to the
  environment, and $\sigma$ would then correspond to the intensity of
  environmental fluctuations. It is yet unclear what interpretation to ascribe
  to $\lambda$ (qualitatively some measure of news coupled with price changes),
  and the question of whether it is an observable remains open. Therefore, we
  estimate the drift and diffusion term and examine their behavior as $\Sigma$
  varies.
\bibitem{fractional}
Since the number of trades $N$ is distributed as a
  power-law~\protect\cite{Plerou00}, or equivalently the time interval between
  trades is distributed as a power-law (with an exponential truncation), the
  stochastic differential equations should be more precisely
  fractional~\protect\cite{Halperin77,Mainardi00}.
\bibitem{notediff}
Our goal is to understand the change in behavior of
$P_{\Omega}(\Omega)$ with $\sigma$. Our results hold under the
assumption that $\sigma$ changes much slower than the scale at which
$\Omega$ changes, so that a stationary $P(\Omega)$ is reached.
\bibitem{noteSigma-sigma}
In our problem, $\sigma$ is not known; however, $\Sigma$ is known, and it
  monotonically depends on $\sigma$. Thus, instead of studying $V
  (\Omega)$ for different $\sigma$, we examine $V(\Omega)$ for different
  $\Sigma$. We note that although $\Sigma$ depends monotonically on $\sigma$,
  the converse need not be true. However, it is convenient to assume that
  changes of $\Sigma$ largely reflect changes in $\sigma$.
\bibitem{note-additivenoise}
Although, we consider only multiplicative noise in our description, in general
  there could be additive noise present as well.
\end{references}
\end{document}